\documentclass[reprint,twocolumn,notitlepage,nofootinbib,showpacs,preprintnumbers,superscriptaddress]{revtex4-1}
\usepackage[T1]{fontenc}
\usepackage[utf8]{inputenc}
\usepackage{bm}
\usepackage{amsmath}
\usepackage{amssymb}
\usepackage{esint}
\usepackage{color}
\usepackage{graphicx}% Include figure files
\PassOptionsToPackage{normalem}{ulem}
\usepackage{ulem}

\usepackage{times}
\usepackage{hyperref}
\usepackage{amsthm,bm,mathrsfs,stmaryrd,amsmath}
\newcommand{\be}{\begin{equation}}
\newcommand{\ee}{\end{equation}}
\newcommand{\ba}{\begin{eqnarray}}
\newcommand{\ea}{\end{eqnarray}}

\def\pa{\partial}
\def\a{\alpha}

\def\d{\delta}

\def\e{\epsilon}

\def\m{\mu}
\def\n{\nu}

\def\r{\rho}

\def\s{\sigma}

\def\o{\omega}

\begin{document}

\title{A unifying framework for ghost-free Lorentz-invariant Lagrangian field theories}

\author{Wenliang LI}
\email{lii.wenliang@gmail.com}
\affiliation{APC, Universit\'e Paris 7, CNRS/IN2P3, CEA/IRFU, Obs. de Paris, Sorbonne Paris Cit\'e, B\^atiment Condorcet, 
10 rue Alice Domon et L\'eonie Duquet, F-75205 Paris Cedex 13, France 
({UMR 7164 du CNRS})}

\begin{abstract}
We propose a framework for Lorentz-invariant Lagrangian field theories 
where Ostrogradsky's scalar ghosts could be absent. 
A key ingredient is the generalized Kronecker delta. 
The general Lagrangians are reformulated in the language of differential forms. 
The absence of higher order equations of motion for the scalar modes 
stems from the basic fact that every exact form is closed. 
The well-established Lagrangian theories for spin-0, spin-1, p-form, spin-2 fields have 
natural formulations in this framework. 
We also propose novel building blocks for Lagrangian field theories. 
Some of them are novel nonlinear derivative terms for spin-2 fields. 
It is nontrivial that Ostrogradsky's scalar ghosts are absent in these fully nonlinear theories. 
\end{abstract}

\maketitle

\section{Introduction}
The problem of ghost-like degrees of freedom are encountered in 
the construction of theories with large numbers of spacetime indices, 
either from high spin fields or high order derivatives. 
High spin fields are dangerous because 
some tensor indices become derivative indices 
for the longitudinal modes, 
while high derivative Lagrangians are dangerous due to Ostrogradsky's instability. 

A host of ghost-free theories were carefully constructed, 
resulting in the emergence of a common pattern.
The linear theory of ghost-free massive gravity requires the Fierz-Pauli tuning \cite{Fierz:1939ix} , 
where the indices of mass terms are contracted antisymmetrically. 
When Lovelock studied the most general metric theories with second order equations of motion \cite{Lovelock:1971yv}, 
the ghost-free combinations turned out to be antisymmetric products of Riemann curvature tensors.
The antisymmetric structure appeared, again and again, 
in the high derivative scalar theories free of Ostrogradsky's ghost \cite{Nicolis:2008in}, 
nonlinear potential terms for massive spin-two fields \cite{deRham:2010kj}, etc.
\cite{vector-galileon, Deffayet:2010zh, Horndeski:1974wa} 

This general pattern also applies to conventional theories. 
Both the Maxwell action for massless spin-1 fields and 
the linearized Einstein-Hilbert action for massless spin-2 fields 
are two-derivative quadratic actions with indices contracted antisymmetrically. 
Antisymmetrization seems to be a universal element in ghost-free Lorentz-invariant Lagrangians. 

It is tempting to develop a unifying framework 
for local, ghost-free, Lorentz-invariant, Lagrangian field theories, 
with antisymmetrization being a key ingredient. 
It is clearly not an easy task to keep track of all the degrees of freedom in full generality  
and to make sure they are all free of ghost-like instability. 
As a first step, we will focus on the scalar modes. 
The absence of scalar ghosts is a necessity for completely ghost-free models. 
Therefore, ghost-free theories belong to a subset of scalar-ghost-free models 
and can be covered in this work. 

In this work, we propose that the Lagrangians should be some differential forms. 
This may seem like a trivial statement as the volume element is itself a differential form and 
a Lagrangian, as the integrand of an action integral over spacetime, 
is defined as the product of a scalar function and the volume form. 

The refinement in our proposal is that 
the Lagrangians should be the wedge products of 
geometric forms and matter forms
\be
\mathcal L=\sum f\, \o_1 \wedge \dots \wedge\o_n, 
\ee 
where the precise meanings of these differential forms $\o_k$ are discussed later. 
In the simplest case, $\o_k=E$ are the same vielbein one-form. 
Their wedge product gives the volume form, 
which appears in general relativity as the cosmological constant term. 
In general situations, $\o_k$ could be the curvature two-forms and 
some exact forms constructed from matter fields. 
The use of these elaborate building blocks can help us to understand 
why higher derivative terms are forbidden from appearing in equations of motion: 
the absence of ghost-like degrees of freedom becomes 
a consequence of a basic property of exterior derivative
\be
d^{\,2}=0,
\ee
whose geometric interpretation by Stokes's theorem is the boundary of a boundary vanishes. 

\section{Ostrogradsky's ghosts}
In this section, let us explain the ghost problem in our discussions. 
Ostrogradsky's theorem states that the energy of a Lagrangian theory 
with higher order time derivative terms
is not bounded from below 
because the Hamiltonian will be linear in a conjugate momentum \cite{Woodard:2006nt}. 
The additional negative energy modes are ghost-like degrees of freedom 
and their presence is due to the fact that 
equations of motion are of higher than second order. 
A loophole in Ostrogradsky's proof is the assumption that the Lagrangian is non-degenerate. 
In other words, if the Euler-Lagrange equations remain second order, 
a higher derivative Lagrangian theory could be healthy and 
no additional problematic degree of freedom is propagating. 

The absence of higher order time derivative terms 
in the equations of motion is not a sufficient condition for healthy models. 
For example, a Hamiltonian could be unbounded due to a wrong-sign kinetic term or potential term. 

A more subtle point is that, in high spin field theories, 
the Hamiltonians can be unbounded from below, 
even if Ostrogradsky's ghosts are absent and 
the Lagrangian does not contain any wrong sign. 
An important example is the linearized Einstein-Hilbert term, 
whose Hamiltonian is unbounded due to the opposite signs of two momentum squared terms\footnote{The linear theories of massless higher spin fields may share the same issue. 
An explicit example can be found in \cite{Campoleoni:2016uwr} , 
where the Fronsdal action \cite{Fronsdal:1978rb}  for a spin-3 field is rewritten in Hamiltonian form. }
\footnote{In the Hamiltonian form of the Einstein-Hilbert action, 
the momentum squared terms are included in the Hamiltonian constraint and 
the Hamiltonian simply vanishes on the constraint surface. }. 

Having these subtleties in mind, 
we would like to confine our attention to the absence of Ostrogradsky's ghosts 
in the sense that no additional degree of freedom is propagating 
and the equations of motion are at most of second order. 
Whether the Hamiltonians are bounded from below in the end is beyond this work. 
To further simplify our discussion, we will mainly concentrate on the scalar modes 
and require the absence of Ostrogradsky's scalar ghosts. 

\section{Lagrangians free of Ostrogradsky's scalar ghosts}
Now we want to discuss general Lagrangians that could be free of Ostrogradsky's scalar ghosts. 
We will explain how the generalized Kronecker delta arises
as a result of second order equations of motion together with Lorentz invariance. 
Then we will derive the general form of ghost-free Lorentz-invariant Lagrangians in the language of tensors. 

By Lorentz-invariance, we mean that Minkowski vacuum is a solution of the models 
and, after we expand the Lagrangians around this background solution, 
the field contents transform properly under Lorentz transformations.
The action should be invariant under these global symmetry transformations. 
So the theories do not distinguish among time and different space indices up some signs
\footnote{Some of our results can be generalized to other maximally symmetric vacua, 
i.e. de-Sitter space and Anti de-Sitter space. For example, the cosmological constant term mentioned in the introduction can lead to dS or AdS background solutions. 
Background-independent ghost-free theories should reduce to 
Lorentz-invariant ghost-free theories if the limit is not singular, 
so they could be obtained by proper generalizations of 
a subset of Lorentz-invariant ghost-free theories. }. 
This definition of Lorentz-invariance applies to gravitational theories 
when the Minkowski metric is a solution. 

A Lagrangian constructed from zeroth and first order terms will not lead to 
apparent higher order equations of motion. 
Let us consider a Lagrangian with harmless zeroth order terms and dangerous second order derivative terms 
\be
\mathcal L\sim \phi\dots\phi\,\pa\pa \phi\dots \pa\pa \phi,
\ee
where $\phi$ indicates dynamical fields and they can have tensor indices. 
Without $\phi\dots\phi$, the equations of motion will be of 
higher than second order or simply vanish.
Higher than second order derivative terms are not considered 
because they usually lead to terms of at least the same order 
in the equations of motion after varying the $\phi\dots\phi$ part with respect to $\phi$.  
We postpone the inclusion of first order derivative terms to later discussions. 

Now we examine the variations of the product of two second order terms
\be
\delta(\pa_{a_1}\pa_{a_2} \phi\, \pa_{b_1}\pa_{b_2}\phi\dots)\rightarrow 
\left(
\pa_{a_1} \pa_{b_1}\pa_{b_2}\phi\,\pa_{a_2}\dots
+\,\dots
\right),
\ee
where we concentrate on a third order term on the right hand side. 
There are fourth order derivative terms as well, but the spirit manifests itself already 
in the third order terms. 
Since derivatives commute with each others, third order derivative terms with the same indices but different orders are equivalent. 
The coefficient of a third order derivative term is
\begin{align}
C^{\m,\n\r}\pa_\m\pa_\n\pa_\r\phi=
\Big(C^{a,bc}+C^{a,cb}+C^{b,ac}+C^{b,ca}
\nonumber\\\,
+C^{c,ab}+C^{c,ba}
\Big)\pa_{a} \pa_{b}\pa_{c}\phi.
\label{cub}
\end{align}
To have a vanishing coefficient
\footnote{We assume all higher order derivative terms should be absent, 
which seems to be stronger than the absence of higher order time-derivative terms. 
For example, $\pa_i \pa_0\pa_0\phi$ is of second order in time derivatives, 
but it comes from a third order term $\pa_\m\pa_\n\pa_\r\phi$. 
If $\pa_0^3\phi$ is eliminated by some special tensor structure, 
then the third order term $\pa_\m\pa_\n\pa_\r\phi$ is allowed.}, 
there are two simple choices: 
either $(\m,\n)$ or $(\m,\r)$ are antisymmetrized. 
A more detailed derivation is given in section II of \cite{Li:2015fxa}. 
For the second cubic derivative term, we impose the same requirement, 
then we have two ansatzes to obtain second order equations of motion: 
\begin{itemize}
\item
$(a_1,\,b_1)$ and $(a_2,\,b_2)$ are two sets of antisymmetrized indices;
\item
$(a_1,\,b_2)$ and $(a_2,\,b_1)$ are two sets of antisymmetrized indices.
\end{itemize}
The same requirements for other second order derivative terms lead to two chains of antisymmetrized indices for the derivative indices of the second order terms. 
These tensor structures correspond to Young diagrams with two columns.

Let us now consider first order derivative term $\pa\phi$. 
At first sight, varying the first order term will lead to a third order term
\be
\delta(\pa\phi)\,\pa\pa\phi\rightarrow -\pa\pa\pa\phi,
\ee
but, in the case of single scalar field, 
this term is cancelled by varying the corresponding second order term
\be
\pa\phi\,\delta(\pa\pa\phi)\rightarrow \pa\pa\pa\phi,
\ee
so $\pa\phi$ is harmless. 
The difference in signs is due to the different numbers of integration by parts in the two cases. 

Now we discuss the tensor indices in the dynamical fields. Let us concentrate on the longitudinal scalar mode 
\be
\phi_{\m_1  \dots\m_k}\sim\pa_{\m_1}\dots\pa_{\m_k} \Phi,
\label{lon}
\ee
which is the most dangerous due to the large number of derivatives in front of it. 
In other words, we focus on the longitudinal part of the Helmholtz decomposition of a tensor field. 
Substituting eq.\eqref{lon} into the equations of motion,  
second order derivative terms become of higher order. 
To eliminate these higher order terms, 
we include the tensor field indices to the chains of antisymmetrized indices. 

Extending the analysis of first order derivative terms, single-index terms 
(spin-1 fields and first derivative of spin-0 fields) 
are not dangerous in certain situations (see \cite{Li:2015fxa} for more details). 
We can also include their indices to the antisymmetric chains, but that is not necessary. 
In contrast, if we have fields with spins higher than one 
or consider interaction among multiple fields, 
more care should be taken as their first derivative terms are usually dangerous. 

From the two chains of antisymmetrized indices, 
we then construct general scalar-ghost-free Lagrangians 
\be
\mathcal L=\sum d^D x\,f(\phi,\pa\phi)\,\d_{\n_1\n_2\dots}^{\m_1\m_2 \dots}\prod \o_{\m\dots}^{\n\dots}\quad,
\label{gen-form}
\ee
where $f$ are scalar functions of scalar fields (and single-index terms in some cases), 
$\d_{\n\dots}^{\m \dots}$ is the generalized Kronecker delta 
defined as the product of two Levi-Civita symbols with contracted indices
\be
\delta_{\n_1\n_2\dots\n_k}^{\m_1\m_2\dots\m_k}=\epsilon_{\n_1\n_2\dots\n_D}\, \epsilon^{\m_1\m_2\dots\m_D}\delta_{\m_{k+1}}^{\n_{k+1}}\dots \delta_{\m_D}^{\n_D}\quad, 
\label{gen-Kron}
\ee
and 
$\o$'s denote dynamical fields or their derivative terms 
\be
 \o_{\m\dots}^{\n\dots}=
 \phi_{\m\dots}^{\n\dots},\quad
  (\pa\phi)_{\m\dots}^{\n\dots},\quad
 (\pa\pa\phi)_{\m\dots}^{\n\dots}\quad.
 \label{dphi}
\ee
In eq.\eqref{dphi} , 
the number of derivatives acting on the dynamical fields $\phi$ could range from 0 to 2. 
The two chains of antisymmetrized indices are denoted by $\m_i$ and $\nu_i$. 
For symmetric fields, $\o$'s have at most four indices: two for derivatives and two for tensor fields. 
We may also consider p-form fields or mixed-symmetry fields with two columns 
\footnote{However, for bosonic, symmetric, massless spin-s fields with $s>2$, 
the quadratic action will vanish if it is an antisymmetric product in the form of eq.\eqref{gen-form}. 
In fact, the Fronsdal action \cite{Fronsdal:1978rb} for linear massless spin-s fields with $s>2$
is not an antisymmetric product. 
The corresponding equations of motion do propagate a correct number of degrees of freedom, 
but, unlike the spin-2 case, the longitudinal mode $\phi_{\m_1  \dots\m_k}\sim\pa_{\m_1}\dots\pa_{\m_k} \Phi$ is still dynamical and 
the equations for $\Phi$ is of higher than second order. 
The presence of the longitudinal scalar mode $\Phi$ is 
due to the subtle traceless constraint on 
the gauge parameter. 
In this sense, the linear theories of higher spin fields are plagued by 
Ostrogradsky's scalar ghosts. 
}. 
For second derivative terms, 
the derivative indices should be in different chains of antisymmetrized indices. 
In addition, the indices of a spin-2 field should be in different chains. 

To obtain the generalized Kronecker delta, 
we assume the indices in the two chains are contracted with each others. 
If free indices are allowed, they will be contracted with those of the harmless terms 
with single indices. 
This will effectively introduce the indices of the single-index terms to the antisymmetric chains. 
Therefore, the Lagrangians are again in the general form \eqref{gen-form}. 

\section{Differential forms}\label{diff}
We want to reformulate the general scalar-ghost-free Lagrangians \eqref{gen-form} 
in the language of differential forms.
This is motivated by the crucial importance of the index antisymmetrization, 
which indicates the language of differential forms may be more natural. 

Here is another motivation. 
Lovelock's theory was reformulated by Zumino in terms of vielbein 
and spin-connection in \cite{Zumino:1985dp} . 
The absence of high derivative terms in the equations of motion is due to the Bianchi identities. 
These identities are from a more basic identity of the exterior derivative: $d^{\,2}=0$.
We want to generalize this statement to other ghost-free Lagrangian field theories in this framework.

In terms of differential forms, the general scalar-ghost-free Lagrangians \eqref{gen-form} correspond to D-forms 
\footnote{See \cite{Ezquiaga:2016nqo} also for detailed discussions of 
scalar-tensors theories in the Language of differential forms. }
\be
\mathcal L= \sum f \,\o_1 \wedge \dots \wedge \o_k\wedge\bar e\wedge\dots \wedge \bar e,
\label{gen-form-diff}
\ee
where the differential forms $\o_1,\o_2,\dots$ and $\bar e$ are defined as
\be
\o^{\n_1\n_2\dots}=\o^{\n_1\n_2\dots}_{\m_1\dots\m_k} dx^{\m_1}\wedge\dots\wedge dx^{\m_k},
\ee 
\be
\bar e^\n=\bar e_\m{}^\n dx^\m,\quad \bar e_\m{}^\n=\d_\m{}^\n,\label{eta}
\ee
and $D$ is the number of spacetime dimensions.
Note that the upper indices $\n_k$ in eq.\eqref{gen-form-diff} 
can be interpreted as the internal indices of local Lorentz frames 
and they are contracted with a Levi-Civita symbol $\e_{\n_1\dots\n_D}$. 
We interpret the frame field $\bar e$ in eq.\eqref{eta} as the Minkowski vielbein 
\footnote{Here we consider a simple gauge for the local Lorentz frames 
where the Minkowski vielbein is a constant given by eq.\eqref{eta}. 
In a general gauge, the components of a Minkowski vielbein $\bar e$ are spacetime functions 
and they are related to eq.\eqref{eta} by internal Lorentz transformations. 
Then the $\m$-derivatives $\pa_\m$ should be supplemented by spin connections 
and become covariant derivatives $\bar D_\m$.  
Although the Christoffel symbols vanish for flat spacetime, 
the spin-connections are non-zero due to the spacetime-dependence of the Minkowski vielbein. 
On the other hand, the covariant $\n$-derivatives are related to 
the covariant $\m$-derivatives by $\bar e_{\m}{}^\n$ 
\be
\pa^\n\rightarrow \bar e_{\m}{}^\n \bar D^\m.
\ee
Accordingly, tensors with upper indices are defined through the Minkowski vielbein
\be
T^{\n_1\n_2\dots}=T^{\m_1\m_2\dots} \bar e_{\m_1}{}^{\n_1}\bar  e_{\m_2}{}^{\n_2},
\ee
where the spacetime indices $\m_k$ are raised by 
the inverse Minkowski metric $\eta^{\m_1\m_2}$. 
}, 
and they are from the Kronecker deltas in eq.\eqref{gen-Kron}. 
Terms with $\m$-derivatives are exact forms 
\footnote{But the $\n$-derivative terms are not necessarily exact forms in this formulation. 
For example, $\pa^\n \phi$ is invariant under coordinate transformations, 
but transforms as a Lorentz vector under local Lorentz transformations.}.
For instance, the $\m$-derivative of a spin-1 field can be an exact one-form field 
\be
\o^\n=\pa_\m A^\n\, dx^\m=d\, A^\n,
\ee
or an exact two-form field (field strength)
\be
\o=\pa_{\m_1} A_{\m_2}\, dx^{\m_1}\wedge dx^{\m_2}=d\, A. 
\ee
Varying the actions with respect to $A^\n$ or $A$, 
the exterior derivative $d$ will move to other terms after an integration by parts. 
When it acts on a second derivative term,  
the dangerous cubic order term vanishes thanks to
\be
d^{\,2}=0.
\ee 
We can see the absence of higher order equations of motion originates in 
the basic fact that every exact form is closed, 
as anticipated in the case of Lovelock theory. 

Note that the formulations in the language of differential forms have a duality 
as the theories are invariant under the exchange between $\m$- and $\n$- indices. 
The above discussion applies to the $\n$-derivatives as well. 
This duality generalizes the Hodge duality in Maxwell's theory. 

\section{Examples}
In this section, let us discuss some concrete examples of the scalar-ghost-free Lagrangians.
\subsection{Single spin-0 fields}
In the case of single scalar field, 
the ghost-free terms in eq.\eqref{gen-form} become
\be
\label{sca-1}
\mathcal L_{1,n}^{(s=0)}=f(\phi,\pa^\r\phi\,\pa_\r\phi)\,\d_{\n_1\dots\n_n}^{\m_1 \dots\m_n}\,\prod_{k=1}^n \pa_{\m_k}\pa^{\n_k}\phi
\ee
and
\be
\label{sca-2}
\mathcal L_{2,n}^{(s=0)}=f(\phi,\pa^\r\phi\,\pa_\r\phi)\,\d_{\n_1\dots\n_n}^{\m_1 \dots\m_n}\,
\pa_{\m_1}\phi\,\pa^{\n_1}\phi\prod_{k=2}^{n} \pa_{\m_k}\pa^{\n_k}\phi,
\ee
where $(s=0)$ indicates only one spin-0 field is under consideration and 
$n$ labels the numbers of derivatives. 
The ghost-free terms \eqref{sca-1} have no antisymmetrized single-index term, while the second ghost-free terms \eqref{sca-2} have two such terms. 
If we consider more single-index terms, the ghost-free term will vanish due to antisymmetrization. 
The two kinds of ghost-free terms \eqref{sca-1} and \eqref{sca-2} are related by using the properties of generalized Kronecker delta and integrating by parts. 
They are two equivalent formulations of the Galileons, 
which are the most general ghost-free, Lorentz-invariant, single scalar field theories. 
In this case, we do not have the freedom to construct novel models due to the limited number of indices. 

\subsection{Single spin-1 fields}
The general scalar-ghost-free terms \eqref{gen-form} for single vector fields read 
\begin{align}
\mathcal L_1^{(s=1)}=&\, f( A_\mu A^\mu)\delta^{\m\dots}_{\nu
\dots}
[\pa_{\m}A_{\m}]
[\pa^{\n}A^\n]
\nonumber\\&\qquad
[\pa_{\m}A^{\n}]
[\pa^{\n}A_\m]
[\pa_{\m}\pa^{\n}A_{\m}]
[\pa_{\m}\pa^{\n}A^{\n}],
\label{vec-1}
\end{align}

\begin{align}
\mathcal L_2^{(s=1)}=&\, f( A_\mu A^\mu)\delta^{\m\dots}_{\nu
\dots}
A_\m
[\pa_{\m}A_{\m}]
[\pa^{\n}A^\n]
\nonumber\\&\qquad
[\pa_{\m}A^{\n}]
[\pa^{\n}A_\m]
[\pa_{\m}\pa^{\n}A_{\m}]
[\pa_{\m}\pa^{\n}A^{\n}],
\label{vec-2}
\end{align}
and
\begin{align}
\mathcal L_3^{(s=1)}=&\, f( A_\mu A^\mu)\delta^{\m\dots}_{\nu
\dots}
A_\m A^\n
[\pa_{\m}A^{\n}]
[\pa^{\n}A_\m]
\nonumber\\&\qquad
[\pa_{\m}A_{\m}]
[\pa^{\n}A^\n]
[\pa_{\m}\pa^{\n}A_{\m}]
[\pa_{\m}\pa^{\n}A^{\n}],
\label{vec-3}
\end{align}
where $[X]$ denotes a product of $X$'s
\be
[X]=\prod X.
\ee
To simplify the notation, the $i$ indices in $\m_i,\,\n_i$ are not written explicitly. 
Note that eqs.(\ref{vec-1}-\ref{vec-3}) are not completely independent terms. 

For U(1) gauge invariant theories, a no-go theorem was established in \cite{Deffayet:2013tca}. 
By breaking the gauge invariance, non-trivial vector Galileons were constructed in \cite{vector-galileon}. 
They are special cases of \eqref{vec-1} and \eqref{vec-3}. 
Eq. \eqref{vec-2} and second derivative term $\pa_\m\pa^\n A_\r$ were not discussed in \cite{vector-galileon}.
Therefore, 
eq. \eqref{vec-2} and eqs. (\ref{vec-1}, \ref{vec-3}) involving $\pa_\m\pa^\n A_\r$ are new terms 
if they are not trivially related to the previous results by integrating by parts.

Along this direction, the p-form Galileons in \cite{Deffayet:2010zh} can be generalized to include 
zeroth order derivative terms and
first order derivative terms that are different from the field strengths. 
The first order terms are not necessary the field strengths 
because we have two chains of antisymmetrized indices. 
In the case of differential forms, we should then consider the Hodge decomposition. 
Due to the antisymmetric nature of p-form fields, we can substitute at most one tensor index with a derivative index. 
Then the most dangerous mode in a p-form is its exact part 
and all the tensor indices should be absorbed in one of the antisymmetric chains. 
The function $f$ could depend on p-form fields if there is no second derivative acting on a p-form field. 
To summarize the p-form generalizations, 
both the derivative indices and the p-form indices should be 
included into the antisymmetric chains, 
assuming that the Lagrangian can written as a wedge product of the p-form and its exterior derivative. 

In four dimensions, the number of inequivalent theories are significantly reduced 
because the length of a antisymmetric chain cannot be larger than the number of spacetime indices.
Below we examine a novel theory constructed from second derivative terms. 
Its Lagrangian reads
\be
\mathcal L_{\text{ex}}^{(s=1)}=\delta^{\m_1\m_2\m_3\m_4}_{\nu_1\n_2\n_3\n_4}
A_{\m_1}A^{\n_1}
\pa_{\m_2}\pa^{\n_2}A_{\m_4}
\pa_{\m_3}\pa^{\n_3}A^{\n_4}.
\ee

The zero component of the conjugate momentum is
\be\label{spin-1-momenta}
(\pi_{\text{ex}}^{(s=1)})^0=\frac {\pa\mathcal L_{\text{ex}}^{(1)}}{\partial \dot A_0}
=-2(\pa_{i}\pa^{[i}A_{j})\,\pa^{j}(A_{k}A^{k]}).
\ee
The time derivative terms in $\pi^0$ cancel out, so eq. \eqref{spin-1-momenta} is a primary constraint. 
To preserve this constraint in time, we obtain a secondary constraint by computing the Poisson bracket of the Hamiltonian and the primary constraint. 
These two constraint equations eliminate 
the longitudinal ghost-like degree of freedom 
in this higher derivative theory. 
One can substitute $A_\m$ with $A^T_\m+\pa_\m \Phi$ and verify that
the equations of motion are still of second order. 

By construction, the absence of time derivative terms in $\pi^0$ is 
a universal feature of these theories of single spin-1 fields. 
The reason is that the 0 indices in the two antisymmetric chains are already used in $\dot A_0$.

\subsection{Single spin-2 fields}
The general ghost-free Lagrangians \eqref{gen-form} for single dynamical spin-2 fields are
\be
\mathcal L^{(s=2)}=\sum
\delta^{\m\dots}_{\nu
\dots}
[{h_\m}^\n]
[\pa_{\m}{h_\m}^\n]
[\pa^{\n}{h_\m}^\n]
[\pa_{\m}\pa^{\n}{h_\m}^\n]. 
\ee
The scalar functions $f$ can only be constants as spin-2 fields have at least two indices 
and they should live in the antisymmetric chains. 
We do not have the ambiguity of single-index terms anymore. 
By integrating by parts, the general scalar-ghost-free terms are
\be
\mathcal L_{i,j}^{(s=2)}=
\delta^{\m_1\dots\m_{2i+j}}_{\n_1
\dots\n_{2i+j}}
\prod_{k=1}^{i}\pa_{\m}\pa^{\n}{h_{\m}}^{\n}
\,\prod_{k=1}^{j}{h_{\m}}^{\n}. 
\label{gra}
\ee
We can see the spin-2 theories are more constrained than the spin-1 theories due to the large numbers of indices. 
When $j=0$, the Lagrangian is a total derivative, so we require $j>0$. 
For $j=1$, the Lagrangians $\mathcal L_{i,1}^{(2)}$ are the leading nontrivial terms of perturbative Lovelock terms. 
The linearized Einstein-Hilbert term corresponds to the case $i=j=1$. 
The Fierz-Pauli term is $\mathcal L_{0,2}^{(2)}$. 
The other zero-derivative $(i=0)$ interaction terms $\mathcal L_{0,j}^{(2)}$ can be thought of as the perturbative terms 
from the dRGT potentials for (massive) spin-2 fields. 
In 4D, an interesting two-derivative term was discovered in \cite{Folkerts:2011ev} 
\begin{equation}
\mathcal L_{1,2}^{(s=2)}=h_\m{}^{[\m} h_{\n}{}^\n \pa_\r\pa^\r h_\s{}^{\s]},
\end{equation}
which is different from the cubic perturbative terms from the Einstein-Hilbert action. 
There are more new derivative terms in higher dimensions, 
and they can be understood as natural generalizations of perturbative Lovelock 
and dRGT terms \cite{Hinterbichler:2013eza}. 

It is known that the linear theory of massless spin-2 field can be extended 
to a nonlinear theory with the same number of dynamical degrees of freedom. 
The result is the Einstein-Hilbert action. 
An interesting question is: 
Do the new scalar-ghost-free terms in eq.\eqref{gra} admit nonlinear extensions? 
In other words, we want to interpret the spin-2 field $h_{\m\n}$ as a perturbation around 
the Minkowski metric
\be
g_{\m\n}=\eta_{\m\n}+h_{\m\n}.
\ee 
And the full actions should be functionals of the basic dynamical variable $g_{\m\n}$,  
rather than the perturbation $h_{\m\n}$, and 
eq.\eqref{gra} are the leading perturbative terms of the full Lagrangians. 
It was proposed in \cite{Hinterbichler:2013eza} that 
$\mathcal L_{1,2}^{(2)}$ corresponds to the perturbative term of 
a new nonlinear kinetic term for 4D massive spin-2 fields. 
Nonlinear extensions of $\mathcal L_{1,2}^{(2)}$ 
and other new derivative terms were constructed in \cite{Kimura:2013ika}. 
However, they were shown to propagate the Boulware-Deser ghost \cite{Kimura:2013ika}\cite{deRham:2013tfa}, 
which is known as the Ostrogradsky's scalar ghost 
in generic nonlinear theories of massive spin-2 fields. 

In principle, one can construct the nonlinear theory order by order in fields. 
This powerful perturbative procedure slows down at high orders and 
to obtain the complete nonlinear theory might require endless work. 

Below we use some geometric intuitions to obtain the general nonlinear ghost-free gravitational theories. 
In the language of differential forms, vielbeins are more natural building blocks than metrics. 
The second derivative of a vielbein does not make much sense from the perspective of geometry. 
A more geometric candidate for the nonlinear two-derivative term is the curvature two-form. 
Then we arrive at natural nonlinear extensions of eq.\eqref{gra}
\be
\tilde{\mathcal L}^{(s=2)}_{i,j}=R\wedge\dots \wedge R\wedge E\wedge\dots\wedge E\wedge \bar e\wedge \dots \wedge \bar e,\label{non-s2}
\ee
which are the wedge products of the curvature two-form $R(E)$, the dynamical vielbein $E$ 
and the fixed Minkowski vielbein $\bar e$. 
\footnote{After the first edition of this draft appeared in arXiv, 
we were soon informed that the proposal \eqref{non-s2} was studied before 
by K. Hinterbichler and R. A. Rosen. }
These nonlinear extensions are consistent with 
the interpretation in Sec. \ref{diff} that 
$\n$-indices in eq.\eqref{gen-form-diff} are local Lorentz indices. 
The subscripts $(i,j)$ indicate the numbers of the curvature two-form and 
the dynamical vielbein in the wedge products. 
There are $(D-2i-j)$ Minkowski vielbein $\bar e$ in the wedge products. 
The definition of $\bar e$ is given by eq.\eqref{eta}. 
Spin-connections may be introduced to eq.\eqref{non-s2} as geometric completion of first order terms, 
but we leave this possibility for future investigations. 
Several nonlinear metric theories are recovered as special cases of eq.\eqref{non-s2}. 
The Lovelock terms correspond to $\mathcal L^{(s=2)}_{i,D-2i}$ and 
the Einstein-Hilbert action is $\mathcal L^{(s=2)}_{1,D-2}$.
The dRGT terms in the vielbein reformulation \cite{Hinterbichler:2012cn} are $\mathcal L^{(s=2)}_{0,j}$. 

Eq. \eqref{non-s2} also generates some novel nonlinear theories for single dynamical spin-2 field. 
Unfortunately, when symmetric conditions are imposed, 
they are equivalent to the nonlinear proposals in \cite{Kimura:2013ika}, 
which means Ostrogradsky's scalar ghosts are present. 
We will come back to the nonlinear theories of spin-2 fields in the Sec. \ref{sec-cov}. 

\subsection{Coupled multiple fields}
The choice is rich for ghost-free interactions of multiple fields. 
An important difference from the single field theories is that 
most of the single-index terms should be in the antisymmetric contraction.  
The interaction terms for multiple fields are the wedge product of the possible terms. 
For example, in the simplest case, the Lagrangians of bi-/ multi-galileons \cite{Padilla:2010de} 
can be reformulated as 
the wedge products of the exact matter forms constructed from different scalar fields
\footnote{In the dual formulation, $\pa^\n \phi^I$ corresponds to an exact form as well.}
\be
\pa_\m\pa^\n \phi^I dx^\m=d(\pa^\n \phi^I),\quad \pa_\m \phi^I dx^\m=d\phi^I. 
\ee

In the abstract expression eq.\eqref{dphi}, 
if we substitute $\phi$ with different fields, 
then we obtain the interaction terms among multiple fields 
which can have different spins.

\section{Covariantization}\label{sec-cov}
A subtle issue concerning gravity is that 
the Minkowski vielbein $\bar e^\n$ should be replaced by a dynamical vielbein $F$ and 
partial derivatives be substituted by covariant derivatives. 
The actions are also supplemented by an Einstein-Hilbert term for the dynamical vielbein $F$.  
This covariantizing procedure is required by the covariance and the universal coupling of gravity. 

The introduction of covariant derivatives may lead to higher order equations of motion. 
Counter-terms were introduced to the action \cite{Deffayet:2009wt} 
in order to cancel the dangerous higher order terms. 
The counter-terms should be proportional to the Riemann curvature tensor 
because they are invisible when the metric is flat. 
An example is Horndeski theory as covariantized galileons 
whose Lagrangians are specific sums of the wedge products of curvature two-form $R_{\m_1\m_2}{}^{\n_1\n_2}$ 
and covariantized matter forms $(D_\m \pa^\n \phi) dx^\m$.
\footnote{Recently, scalar-tensor model beyond Horndeski model was proposed in 
\cite{Gleyzes:2014dya}, 
where the additional counter-terms are not necessary to avoid Ostrogradsky's scalar ghost. 
In some cases, they are related to Horndeski theory by generalized disformal transformations \cite{Zumalacarregui:2013pma}.} 

The same procedure can be applied to spin-1 and spin-2 fields. 
However, when the building blocks $\o$'s in eq.\eqref{dphi} contain more than 2 indices, 
such as $\pa_{\m_1} \pa^{\n_1} A_{\m_2}$,  $\pa_{\m_1} h_{\m_2}{}^{\n_1}$, 
$\pa_{\m_1} \pa^{\n_1}h_{\m_2}{}^{\n_2}$, 
whether consistent covariantization is possible can be a nontrivial question. 
For example, consistent covariantization of the linearized Einstein-Hilbert term $\mathcal L_{1,1}^{(s=2)}$ can lead to the fully nonlinear Einstein-Hilbert term 
which contains a linear term and infinitely many higher order perturbative terms 
around a generic curved spacetime. 

Let us come back to the nonlinear theories of spin-2 fields. 
In eq.\eqref{non-s2}, we have two vielbeins: one is dynamical, 
while the other one is fixed as a reference vielbein. 
The use of Minkowski reference vielbein is partly motivated 
by the dRGT massive gravity \cite{deRham:2010kj} 
in the vielbein formulation \cite{Hinterbichler:2012cn} 
\be
\mathcal L_{dRGT}=
\tilde{\mathcal L}^{(s=2)}_{1,D-2}
+\sum_{i=0}^D \a_i\,\tilde{\mathcal L}^{(s=2)}_{i,D-i},
\ee
where $\a_i$ are the coefficients of the potential terms. 
We can covariantize this model by promoting the fixed vielbein to be dynamical as well, 
then we obtain a nonlinear interacting theory of two spin-2 fields \cite{Hassan:2011zd}.  
The absence of the Boulware-Deser ghost was first established in dRGT massive gravity \cite{Hassan:2011hr}, 
but this property can be extended to the cases of bi-gravity \cite{Hassan:2011zd} 
and even multi-gravity \cite{Hinterbichler:2012cn}. 

In retrospect, dRGT massive gravity can be understood 
as a limit of the bi-gravity models constructed by Hassan and Rosen, 
and so does the fact that the Boulware-Deser ghost is absent in dRGT massive gravity. 
The fixed vielbein has a dynamical origin and is no longer introduced by hand. 
In general, if no dynamical mechanism is accompanied, 
the introduction of a fixed vielbein could be both unnatural and problematic, 
as general covariance is broken explicitly. 
 
The presence of fixed spin-2 fields should not be taken for granted. 
\footnote{It is admitted that usually they can come from the limit 
where the corresponding Planck mass is sent to infinity and the effects of sources are neglected. }
And we should consider a more general setting: 
substitute a new dynamical vielbein $F$ for the Minkowski vielbein $\bar e$ in eq.\eqref{non-s2}
\be
R\wedge\dots \wedge R\wedge E\wedge\dots\wedge E\wedge F\wedge \dots \wedge F.
\label{bi-new-gra}
\ee 
They are the wedge products of the curvature two-form $R(E)$ and 
the dynamical vielbeins $E,\,F$. 
These contain non-standard nonlinear kinetic terms for two spin-2 fields. 
They are investigated in more detail in \cite{gr-kin} and \cite{Li:2015iwc} , 
where the parameter space of interacting theories of two spin-2 fields is further extended. 
\footnote{Higher derivative and multi-field generalizations are also straightforward. }
When we consider at most one curvature two-form in the wedge products, 
4d Weyl gravity and 3d New massive gravity \cite{Bergshoeff:2009hq} can be recovered 
as special combinations of eq.\eqref{bi-new-gra} (see \cite{gr-kin} for more details). 
Note that eq. \eqref{bi-new-gra} also contains new higher derivative terms for interacting spin-2 fields, 
which are bi-gravity generalization of the Lovelock terms. 

The fact that the Boulware-Deser ghost is absent in the new models 
\footnote{However, the price to pay is that the linearized kinetic terms have opposite signs. } 
is a nontrivial support for our general framework, as we are considering fully nonlinear models
\footnote{Another nontrivial aspect is that the absence of Ostrogradsky's scalar ghost does not have a simple, direct  explanation in the Lagrangian formulation using the simple argument $d^2=0$. 
This property can be verified in the Hamiltonian formulation \cite{Li:2015iwc}. }. 
Note that we do not assume the existence of a non-dynamical limit for $F$. 
It turns out to be the case that the non-dynamical limit does not exist. 
The failure of nonlinear extensions eq.\eqref{non-s2} proposed in \cite{Kimura:2013ika} can be understood as a result of the absence of a non-dynamical limit.

\section{Conclusions}
To summarize, 
we propose a general framework for Lorentz-invariant Lagrangian field theories that are free of the Ostrogradsky's scalar ghost. 
The ghost-free terms are reformulated in the language of differential forms.
The differential forms in the wedge products are the matter differential forms 
(which are constructed from spin-0, spin-1, p-form, spin-2 matter fields) 
and the geometric differential forms (which include vielbeins and curvature two-forms). 

New building blocks for scalar-ghost-free models are proposed. 
Ostrogradsky's scalar ghost is absent in eq.\eqref{gen-form} by construction. 
But it is not clear whether all of them can be consistently covariantized or nonlinearly extended. 
In the case of spin-2 fields, we propose fully nonlinear completions for the new perturbative terms. 
They are constructed naturally in the vielbein formulation. 
The absence of the Boulware-Deser ghost in these novel nonlinear derivative terms 
\cite{gr-kin} \cite{Li:2015iwc}  
provides evidence for the effectiveness of the language of differential forms. 

General ghost-free theories for half-integer spin fields are not well-explored. 
They are important because fermionic matter in the standard model of particle physics consists of spin-$\frac 1 2$ particles. 
Spinors are associated with the double covering of the Lorentz group. 
To be compatible with general covariance, 
a finite dimensional spinor lives in the local Lorentz frame 
and its coupling to gravity requires the vielbein formulation. 
The reformulation of our framework in the language of differential forms may be a proper starting point.  
In addition, the novel duality might be elucidated in the spinor formulation. 

The conventional paradigm is that 
we need gauge symmetries to eliminate dangerous ghost-like degrees of freedom. 
The developments \cite{deRham:2010kj,vector-galileon} 
in recent years however indicate that, 
even if gauge symmetries are broken, 
we can still have a correct number of dynamical degrees of freedom. 
What persists in these new models is the antisymmetric structure. 
One of the goals of this work is to show that using antisymmetric structure, 
which is weaker than gauge invariance, 
we are still able to avoid the presence of ghosts, 
at least in the case of Ostrogradsky's scalar ghost.  

\begin{acknowledgments}
I would like to give special thanks to X. Gao for numerous stimulating discussions. 
My thanks also go to F. Nitti for encouragement when the idea of this work 
first appeared in the author's PhD thesis. 
I want to thank E. Babichev, C. Charmousis, E. Kiritsis, J. Mourad, V. Niarchos, K. Noui, R. Saito and D. Steer 
for useful comments or/and discussions. 
I also thank C. de Rham, K. Hinterbichler, A. Matas, 
A. Solomon and A. Tolley for correspondence. 
\end{acknowledgments}

\end{document}